\documentclass[12pt]{article}
\usepackage{amsfonts}
\usepackage{amsmath}
\usepackage{amssymb}
\usepackage{latexsym}
\usepackage[dvips]{graphicx}
\usepackage{color}

\newcommand{\polv}{{\phi}}
\newcommand{\hpolv}{\hat{\phi}}
\newcommand{\hPi}{\hat{\Pi}}
\newcommand{\mmu}{\mu}
\newcommand{\mmuu}{{\bar{\mu}}}
\newcommand{\be}{\begin{equation}}
\newcommand{\ee}{\end{equation}}
\newcommand{\bea}{\begin{eqnarray}}
\newcommand{\eea}{\end{eqnarray}}

\numberwithin{equation}{section}

\begin{document}

\begin{titlepage}
\renewcommand{\thefootnote}{\fnsymbol{footnote}}
\renewcommand{\baselinestretch}{1.3}
\medskip

\begin{center}
{\Large {\bf Polymer quantization of the\\[1ex]
Einstein-Rosen wormhole throat}}
\vspace{1cm} 


\renewcommand{\baselinestretch}{1}
{\bf
Gabor Kunstatter${}^\dagger$, 
Jorma Louko${}^\sharp$
and
Ari Peltola${}^\dagger$
\\}
\vspace*{0.7cm}
{\sl
${}^\dagger$ 
Department of Physics, 
The University of Winnipeg\\
515 Portage Avenue, 
Winnipeg, Manitoba, Canada R3B 2E9\\
{[e-mail: g.kunstatter@uwinnipeg.ca, a.peltola-ra@uwinnipeg.ca]}\\[5pt]
}
{\sl
${}^\sharp$ 
School of Mathematical Sciences,
University of Nottingham\\
Nottingham NG7 2RD, United Kingdom\\
{[e-mail: jorma.louko@nottingham.ac.uk]}\\ [5pt]
}
\vspace{8ex}
{\bf Abstract}
\end{center}
We present a polymer quantization of spherically symmetric 
Einstein gravity in which the polymerized variable is the area of the 
Einstein-Rosen wormhole throat. In the classical polymer theory, the singularity is replaced by a bounce at a radius that depends on the polymerization scale. 
In the polymer quantum theory, we show numerically that the area spectrum is evenly-spaced and in agreement with a Bohr-Sommerfeld semiclassical estimate, and this spectrum is not qualitatively sensitive to issues of factor ordering or boundary conditions except in the lowest few eigenvalues. In the limit of small polymerization scale we recover, within the numerical accuracy, the area spectrum obtained from a Schr\"odinger quantization of the wormhole throat dynamics. The prospects of recovering from the polymer throat theory a full quantum-corrected spacetime are discussed. 
\vfill 
Published in Phys. Rev. D \textbf{81}, 024034 (2010)
\hfill 
Revised December 2009
\end{titlepage}

\section{Introduction}

Polymer representation of quantum mechanics \cite{afw,halvorson} has
recently attracted considerable interest as a way to model loop
quantum gravity in a technically simple setting. It has been applied
to quantum mechanical systems including the harmonic
oscillator~\cite{afw}, Coulomb potential \cite{hlw} and $1/r^2$
potential~\cite{klz}. With these singular potentials it avoids 
the classical singularity in a way that is 
distinct from that in which the singularity is absent in the  
usual Schr\"odinger quantization. When applied to
gravitational mini-superspace models, polymer quantization has
resulted into a variety of scenarios in which a classical singularity
is replaced by a quantum mechanical bounce, both in the cosmological
context \cite{green04,lqc1,lqc2,lqc3,corichi08,ding09} and with black holes
\cite{ashtekar05,modesto06,boehmer07,pullin08,nelson08,pk09}.

In this paper we investigate polymer quantization of spherically
symmetric Einstein gravity in a description in which the classical
dynamical variables are adapted to the evolution of the Einstein-Rosen
wormhole throat \cite{friedman,redmount,Louko:1996md}. As expected
from Birkhoff's theorem and from Hamiltonian analyses of spherically
symmetric Einstein gravity
\cite{Thiemann:1992jj,Kastrup:1993br,Kuchar:1994zk}, the phase space
of the reduced Hamiltonian theory is two-dimensional, and the
configuration variable specifies the size of the wormhole throat as it
evolves from the past singularity through the bifurcation two-sphere
to the final singularity. (For generalizations to charged and rotating
black holes, see~\cite{makela:98}.) The advantage of this model,
compared with a number of recent black hole mini-superspace models
based on a spatially homogeneous foliation
\cite{ashtekar05,modesto06,boehmer07,pullin08,nelson08,pk09}, is that
the reduced Hamiltonian is sufficiently simple to allow not just a
study of the effective classical polymer dynamics but also a numerical
extraction of the mass eigenvalues. Further, the model does not rely
on auxiliary structures that are typically encountered in
mini-superspace models of cosmology and black
holes~\cite{corichi08}. A~disadvantage of the model however is that the
reconstruction of a full quantum-corrected spacetime 
would need further input, including a specification of 
where the spatial hypersurfaces of the
Hamiltonian foliation intersect the horizons.

One piece of input that the polymerized throat dynamics model does require is the choice of the phase space variable that will be polymerized. It has been observed within cosmological mini-superspace models that this choice can significantly affect the polymer dynamics, even in its qualitative properties \cite{lqc1,corichi08}. We show that the choice of the polymerized variable has significant consequences for the throat dynamics already at the semiclassical level, and in particular only certain choices replace the classical singularity by a quantum bounce in the effective polymer dynamics. We then focus on the case where the polymerized variable is the \emph{area\/} of the wormhole throat. This choice is motivated by the proposal of an evenly-spaced area spectrum for quantum black holes, obtained from a wide range of premises; see 
\cite{bekenstein74,bekenstein_mukhanov95,kastrup96,barvinsky96,Louko:1996md,das01,medved02,hod98} for a selection, and the references in \cite{Louko:1996md} for a more extensive list. 
We show that polymerizing the wormhole throat area does lead to a bounce. A perhaps unexpected property however is that for a macroscopic black hole the scale of the bounce can be significantly above the Planck scale, unless the polymerization scale itself is taken to be well below the Planck scale. 

From a mathematical viewpoint, a qualitative difference between our system and the $1/r$ and $1/r^2$ potentials considered in \cite{hlw,klz} is that our polymer Hamiltonian is nonsingular, and there is no need to introduce a ``Thiemann trick'' \cite{Thiemann} modification in the Hamiltonian to regularize any singular terms. The eigenstate recursion relation obtained from our Hamiltonian can contain a singularity, with certain factor orderings, but within the orderings that we consider, this singularity acts merely as a boundary condition in the recursion relation. Reassuringly, we find that the semiclassical regime of the theory is not sensitive to the factor ordering. 

Our main results come from a numerical evaluation of the spectrum. We show that the distribution of the large eigenvalues implies an evenly-spaced area spectrum, and these large eigenvalues are in good agreement with a Bohr-Sommerfeld semiclassical estimate. The choice of the factor ordering has a significant effect only on the lowest few eigenvalues. In the limit of small polymerization scale we recover, within the numerical accuracy, the area spectrum obtained from a Schr\"odinger quantization of the wormhole throat dynamics. We have not attempted to support the numerical results by a rigorous investigation of the spectral properties of the polymer Hamiltonian, but comparison with the corresponding Schr\"odinger Hamiltonian analysis \cite{Louko:1996md} strongly suggests that our polymer Hamiltonian is essentially self-adjoint and has a discrete spectrum. 

The paper is organized as follows. In Sec.\ \ref{sec:class} we recall the main features of the classical throat theory and introduce a canonical chart that is suitable for polymerization. A~conventional Schr\"odinger quantization is performed in Sec.~\ref{sec:schr}. In Sec.\ \ref{sec:poly} we give a brief review of polymer quantization, including the classical limit known as effective polymerization, and the effective polymerization of the throat theory is discussed in Sec.~\ref{sec:eff}. Full polymer quantum theory is analysed in Sec.~\ref{sec:full}, including the numerical results for the mass eigenvalues. 
Section \ref{sec:conc} presents a summary and brief concluding remarks. 

Unless otherwise stated, we use Planck units in which $G=c=\hbar =1$.

\section{Classical throat theory} 
\label{sec:class}

We start with a Hamiltonian system with a 
two-dimensional phase space and the Hamiltonian 
\be 
\label{eq:classH1} 
H = \frac{1}{2}\left(\frac{p^2}{r}+r\right),
\ee
where the configuration variable $r$ takes positive values and $p$ is the conjugate momentum. 
The equations of motion reduce to 
\be 
\label{eq:geodesic} 
\dot{r}^2=\frac{2M}{r}-1,
\ee
where $M$ is the conserved value of $H$ and the overdot denotes 
derivative with respect to the time~$t$. As observed in 
\cite{friedman,redmount}, 
\eqref{eq:geodesic} is the equation of a radial timelike geodesic that passes through the bifurcation two-sphere on a Kruskal manifold of mass~$M$, with $r$ being the area-radius of the two-sphere and $t$ the proper time. It was shown in \cite{Louko:1996md} that the Hamiltonian \eqref{eq:classH1} can be obtained by a Hamiltonian reduction of the spherically symmetric sector of Einstein's theory under suitable boundary conditions: the spacelike hypersurfaces are frozen at the (say) left-hand-side $i^0$ of the Kruskal diagram, they evolve at unit rate with respect to the asymptotic Minkowski time at the right-hand-side~$i^0$, and they intersect a radial geodesic through the bifurcation two-sphere so that $t$ coincides with the proper time on this geodesic. The variable $r$ can then be identified as area-radius of the two-sphere on the distinguished
geodesic. If in addition the spatial slices are chosen to intersect the radial timelike geodesic at the point where the two-sphere radius is minimized, the Hamiltonian \eqref{eq:classH1} can be regarded as the Hamiltonian of the Einstein-Rosen wormhole throat. 

The Hamiltonian system thus describes the proper time evolution of the Einstein-Rosen wormhole throat radius as it expands from zero at the past singularity, reaches the maximum value $2M$ at the bifurcation two-sphere, and collapses back to zero at the future singularity. This means that the spacetime dynamics is described in terms of variables that are in a certain sense confined `inside' the black hole. This is qualitatively similar to the recent studies of quantum black holes in terms of spatially homogeneous slicings of the interior \cite{ashtekar05,modesto06,boehmer07,pullin08,nelson08,pk09}, but there are two significant differences. First, our system is fully reduced, with a true Hamiltonian and no constraints. This is an advantage in the sense that implementation of polymer quantization will be relatively straightforward, and it will be easy to compare the results to those of a conventional Schr\"odinger quantization. Second, the foliation of our system is specified at the infinities and at the wormhole throat, but the foliation is largely arbitrary in the intermediate regions, including the locations where the spacelike hypersurfaces cross the future and past branches of the Killing horizon. This arbitrariness is a disadvantage in the sense that additional input on the foliation seems necessary before one could attempt to reconstruct an entire quantum-corrected spacetime from our quantum theory. 

The purpose of this paper is to examine polymer quantization of the wormhole throat Hamiltonian~\eqref{eq:classH1}. As the results of polymer quantization can depend significantly on the precise choice of the polymerized variables~\cite{lqc2,lqc3}, we shall set up the problem in a way that allows polymerization of an arbitrary positive power of the radius: 
we introduce the new canonical chart $(\polv,\Pi)$, where
{\setlength\arraycolsep{2pt}
\begin{subequations} 
\label{eq:polv}
\bea \phi&=&r^{1/\alpha}, \\ \Pi &=& \alpha p r^{1-1/\alpha},
\eea
\end{subequations}}%
and $\alpha$ is a positive constant. The Hamiltonian takes the form 
\be 
\label{eq:classH2}
H = \frac{\polv^\alpha}{2} 
\left( \frac{\polv^{2-4\alpha} \Pi^2}{\alpha^{2}} + 1 \right) . 
\ee
The case $\alpha=1/2$ is geometrically special in that $\polv$ is then
proportional to the throat area. This case is also special
in that if we had retained geometric units, in which $G=c=1$ but $[\hbar] =
(\text{length})^2$, $\Pi$ would be dimensionless precisely for $\alpha
= 1/2$.

\section{Schr\"odinger quantization} 
\label{sec:schr}

As the classical phase space is finite dimensional, the Hilbert space of a conventional Schr\"odinger quantization can be chosen to be square integrable functions on a finite-dimensional configuration space. We choose the configuration variable to be~$\polv$, and we take the inner product to have a flat measure in~$\polv$:
\be 
(\psi_1, \psi_2) = \int_0^\infty \overline{\psi_1(\polv)}
\psi_2(\polv)\, d\polv. 
\label{eq:ip-flat}
\ee
To quantize the Hamiltonian~\eqref{eq:classH2}, we make the replacement 
$\Pi \mapsto -i d/d\polv$ and choose a symmetric factor ordering. For reasons that will emerge in section~\ref{sec:full}, we consider the family of symmetric orderings in which 
\be
\label{eq:schHordering}
\hat{H} = \frac{1}{2} 
\left( - \alpha^{-2}
\polv^{\beta} 
\frac{d}{d\polv} 
\polv^{2 - 3\alpha -2\beta} 
\frac{d}{d\polv} 
\polv^{\beta} 
+ \polv^{\alpha} \right) , 
\ee 
where the ordering parameter $\beta$ takes real values. 

Writing $\polv = x^{2/(3\alpha)}$ and 
$\psi(\polv) = x^{(1/2) -1/(3\alpha)}\chi(x)$, this Schr\"odinger 
theory is mapped to one in which the inner product is 
\be
(\chi_1, \chi_2)_0 = \int_0^\infty \overline{\chi_1(x)}
\chi_2(x)\, dx , 
\label{eq:ipzero}
\ee 
where we have omitted an overall multiplicative constant, 
and the Hamiltonian is 
\be
\hat{H}_0 = \frac{9}{8} 
\left( 
-  \frac{d^2}{dx^2}  
+ \frac{4x^{2/3}}{9} 
+ \frac{(9\alpha +4\beta-2)(3\alpha +4\beta-2)}{36 \alpha^2 x^2}
\right) . 
\label{eq:Hzero}
\ee 
This is the theory discussed in \cite{Louko:1996md}, 
and the parameter denoted in \cite{Louko:1996md} by $r$ has the value 
$(9\alpha + 4\beta -2)/(6\alpha)$. 
$\hat{H}_0$ is essentially self-adjoint for $\beta \ge \frac12$ and for $\beta
\le \frac12 - 3\alpha$, and for other values of of the parameters 
$\hat{H}_0$ has a $U(1)$ family of
self-adjoint extensions, characterized by a boundary condition at
$x=0$. In all cases the spectrum of $\hat{H}_0$ is discrete, and the
large eigenenergies have the WKB estimate 
\be 
\label{eq:asym1}
E^2_{\mathrm WKB} \sim 2k + B + o(1), 
\ee
where $k$ is an integer and $o(1)$ denotes a term that vanishes
asymptotically at large~$E$. The additive constant $B$ depends on
$\alpha$ and $\beta$ and, for $\frac12 - 3\alpha < \beta < \frac12$, 
also on the self-adjoint extension. For certain choices of the 
self-adjoint extension, the WKB estimate differs
from the true eigenvalues by only a few percent already near the ground state~\cite{makela:04}.

\section{Polymer quantization: an overview} 
\label{sec:poly}

Polymer quantization is a loop quantum gravity motivated, unitarily inequivalent alternative to conventional Schr\"o\-dinger quantization \cite{afw,halvorson}. It incorporates fundamental discreteness to the underlying spatial geometry by confining the quantized Hamiltonian dynamics into a discrete spatial lattice. While the overall philosophy of polymer quantization is deeply connected to loop quantum gravity, polymer quantization is a consistent quantization scheme in its own right, and it is of interest  apply this scheme to a variety of systems.

Consider a quantum theory on the real line~$\mathbb{R}$. The polymer Hilbert space is spanned by the normalizable eigenstates $|\mathrm{x}\rangle$ of the position operator with the inner product
\be
\langle\mathrm{x}'|\mathrm{x}\rangle = \delta_{\mathrm{x}'\!,\mathrm{x}} \, , 
\ee
where the quantity on the right hand side is the Kronecker delta. Any state in the polymer Hilbert space has support on only countably many points, which makes it impossible to define the momentum operator as a differential operator. Instead one defines the actions of two basic operators, the position operator $\hat{\mathrm{x}}$ and the finite translation operator $\hat{U}_\mmu$, as
{\setlength\arraycolsep{2pt}
\begin{subequations}
\bea
\hat{\mathrm{x}}\, |\mathrm{x}\rangle &=& \mathrm{x}\, |\mathrm{x}\rangle ,\\
\hat{U}_\mmu|\mathrm{x}\rangle &=& |\mathrm{x}+\mmu\rangle ,
\eea 
\end{subequations}}%
where $\mmu >0$ is the polymerization scale. In general $\mu$ can be a function of $\mathrm{x}$ but in all that follows it is assumed to be a constant. The momentum operator is then constructed from the translation operator \cite{afw}:
\be 
\label{eq:momentum}
\hat{p}=\frac{1}{2i\mmu}\big( \hat{U}_{\mmu}-\hat{U}_{\mmu}^\dagger\big).
\ee
In the $\mmu\to 0$ limit, \eqref{eq:momentum} reduces to the standard momentum operator $\hat{p}=-i\partial_\mathrm{x}$, thus leading to the usual Schr\"odinger quantized system~\cite{corichi07}. For critical discussions of polymer quantum mechanics, see~\cite{critique}.

The (semi)classical regime of polymer quantum theory is obtained by keeping the polymerization scale $\mmu$ fixed and making the replacement $\hat{U}_\mmu \mapsto e^{i\mu p}$, where $p$ is the continuum momentum. This means making in the classical continuum Hamiltonian the replacement $p \mapsto \sin(\mmu p)/\mmu$. This limit, known as the effective polymer theory or effective polymerization, has proved a useful tool with mini-superspace models of quantum cosmology and black holes \cite{lqc1,lqc2,lqc3,corichi08,ashtekar05,modesto06,boehmer07,pullin08,nelson08,pk09,ding09}.

\section{Effective polymer theory} 
\label{sec:eff}

We now turn to the effective polymerization of the Hamiltonian~\eqref{eq:classH2}. We shall see that the effective theory will restrict the physically interesting values of the parameter~$\alpha$. 

The effective Hamiltonian is obtained from \eqref{eq:classH2} by the replacement $\Pi \mapsto \sin(\mmu \Pi)/\mmu$:
\be
\label{eq:effpolH}
H = \frac{\polv^\alpha}{2} 
\left(
\frac{\polv^{2-4\alpha} \sin^2(\mmu\Pi)}{\alpha^{2} \mmu^2} + 1 \right).
\ee
$H$ takes positive values and is a constant of motion. Denoting the
value of $H$ by~$M$, we see that $\polv$ is bounded above by $\polv
\le {(2M)}^{1/\alpha} =: \polv_+$. Note that this upper bound is at the turning
point of the non-polymerized motion, at the bifurcate two-sphere of
the black hole. 

When $\alpha \ge 2/3$, there are solutions that reach $\polv=0$. 
We shall not consider this case further. 

When $\alpha < 2/3$, the motion is oscillatory, with the outer turning point at $\polv = \polv_+$ and the inner turning point at $\polv = \polv_-$, where $\polv_-$ is the unique solution to 
\be
2M = \frac{\polv^{2-3\alpha}}{\alpha^2 \mmu^2} + \polv^\alpha . 
\ee
The outer turning point is at the turning point of the non-polymerized motion, but the inner turning point is a genuine polymerization effect, replacing the singularity of the non-polymerized theory by a bounce. 
There are three qualitatively different subcases: 
\begin{enumerate}
\item $1/2 < \alpha < 2/3$. 
The asymptotic forms of $\polv_-$ at large and small $M$ are 
\begin{subequations}
\begin{align}
\polv_- &\approx {(2M)}^{1/\alpha} , 
&&M \to \infty, 
\\
\polv_- &\approx {(2M \alpha^2 \mmu^2)}^{1/(2-3\alpha)} , 
&&M \to 0.  
\end{align}
\end{subequations}
\item $\alpha = 1/2$. The inner turning point is
\begin{align}
\polv_- 
= \polv_+ 
\left(
\frac{\mmu^2}{4+\mmu^2} 
\right)^2.
\end{align}
\item $\alpha < 1/2$.
The asymptotic forms of $\polv_-$ at large and small $M$ are
\begin{subequations}
\begin{align}
\polv_- &\approx {(2M \alpha^2 \mmu^2)}^{1/(2-3\alpha)} , 
&&M \to \infty ,  
\\
\polv_- &\approx {(2M)}^{1/\alpha} , 
&&M \to 0.
\end{align}
\end{subequations}
\end{enumerate}

In case 1 we have $\polv_-/\polv_+ \to 1$ as $M \to \infty$. This means that for a macroscopic black hole the bounce happens close to the outer turning point, which seems physically undesirable. In case 2 the ratio $\polv_-/\polv_+$ is independent of~$M$, while in case 3 we have $\polv_- \to \infty$ but $\polv_-/\polv_+ \to 0$ as $M\to\infty$. Cases 2 and 3 seem hence more reasonable, and taking $\mmu$ a few orders below unity will in these cases result into a bounce radius much smaller than the size of the black hole. It is perhaps disconcerting that if $\mu$ is of order unity, the bounce in cases 2 and 3 may occur at a macroscopic scale, and in case 2 even at a scale that is comparable to that of the hole. However, macroscopic polymerization effects have been encountered also in the cosmological context~\cite{green04,lqc1,lqc2,lqc3,corichi08,ding09}, and in the absence of a way to recover from the throat theory a full quantum-corrected spacetime, it is not obvious that these properties would be physically unacceptable. We therefore regard cases 2 and 3 as physically interesting. 

In the rest of the paper we will focus on $\alpha =1/2$ for several reasons. First of all, polymerizing the area is motivated by the number of quantization methods that lead to an evenly-spaced area spectrum of a quantum black hole \cite{bekenstein74,bekenstein_mukhanov95,kastrup96,barvinsky96,Louko:1996md, das01,medved02,hod98}, including Schr\"odinger quantization of the throat theory \cite{Louko:1996md}, as seen in~(\ref{eq:asym1}). One therefore expects the qualitative similarities and differences between polymer and Schr\"odinger spectra to be most easily detectable for $\alpha = 1/2$. We speculate that for $\alpha < 1/2$ the polymer mass spectrum may differ more drastically from the Schr\"odinger spectrum, although we still expect the two to agree in the limit $\mmu \to 0$. 

Second of all, with $\alpha =1/2$ we can compare the numerical results to an analytic evaluation of the large eigenvalues by the Bohr-Sommerfeld semiclassical quantization rule. The Bohr-Sommerfeld rule states that adiabatic invariants of a classical system in periodic motion are quantized in units of Planck's constant~$h$. In our system the canonical variable with periodic time evolution is~$\polv$, whereas $\Pi$ is monotonically decreasing, and the turning points occur when $\Pi$ is an integer multiple of $\pi/(2\mmu)$. The relevant adiabatic invariant is therefore $- \oint \phi \dot \Pi \, dt$, where the integral is evaluated over one period in~$\polv$. Setting $\alpha = 1/2$, we have
\bea
nh &= & - \oint \phi \dot \Pi \, dt 
= 2 \int_{\Pi=0}^{\Pi=\pi/(2\mmu)}\phi(\Pi) \, d\Pi
\nonumber\\
&=& \frac{\mmu^3 M^2}{2}\int^{\pi/2}_0\!\!\frac{d\theta}{{(\sin^2 \! \theta +\mmu^2/4)}^2}\nonumber\\[1ex]
&=& \frac{4\pi M^2\left(2+\mmu^2\right)}{{(4+\mmu^2)}^{3/2}} , 
\label{eq:bs-integral}
\eea
where $n$ is a non-negative integer, we have written $\mmu \Pi = \theta$, and in the last equality we have evaluated the elementary integral. 
Recalling that in our units $h=2\pi$, \eqref{eq:bs-integral} gives for $M^2$ the spectrum 
\begin{subequations} 
\label{eq:BS} 
\begin{align}
&M^2= k(\mmu)n , 
\\
& k(\mmu) := \frac{{(4+\mmu^2)}^{3/2}}{2(2+\mmu^2)} . 
\label{eq:kfunc-BS}
\end{align}
\end{subequations}

The key observation about \eqref{eq:BS} is it that it implies an evenly-spaced area spectrum, with the $\mmu$-dependent spacing factor~$k(\mmu)$. In the continuum limit $\mmu\to0$ we have $k(\mmu) \to 2$, and the spectrum agrees with the leading term in the Schr\"odinger spectrum~(\ref{eq:asym1}). The spacing factor has a minimum at $\mmu =\sqrt{2}$, and for large $\mmu$ it asymptotes to $\mu/2$. We will see in section \ref{sec:semi} that this spectrum is in excellent agreement with our numerical results for the full  polymer quantization except for the lowest few eigenvalues.

\section{Full quantum polymer theory} 
\label{sec:full}

We now proceed to the fully polymerized quantum throat theory. The main aim will be a numerical evaluation of the spectrum. 

For the reasons discussed in Sec.~\ref{sec:eff}, we set $\alpha =1/2$, so that the polymerized variable $\polv$ is proportional to the throat area. The quantum polymer Hamiltonian is obtained from \eqref{eq:classH2} as described in Sec.~\ref{sec:poly}: $\hpolv$ acts on the eigenstates of $\polv$ by multiplication, while 
\begin{subequations}
\begin{align}
\label{eq:hPi} 
& \hPi  := \frac{1}{2i\mmu}\big( \hat{U}_{\mmu}-\hat{U}_{\mmu}^\dagger\big) , 
\\
& \hat{U}_{\mmu} |\polv \rangle = |\polv+\mmu\rangle , 
\end{align}
\end{subequations}
where $\mmu>0$ is the same polymerization scale that appeared in the effective polymer theory of Sec.~\ref{sec:eff}.
In order to deal with negative values of $\polv$ and fractional powers of~$\polv$, we define 
\be 
\label{eq:hpolv} 
\hpolv^{\gamma}|\polv\rangle := 
| \polv|^{\gamma}|\polv\rangle 
\ee
for any real number~$\gamma$. 

The kinetic term in the Hamiltonian presents an issue of factor ordering. We consider two orderings in turn.

\subsection{Nonsingular factor ordering}
\label{sec:nonsing}

In this subsection we order the polymer Hamiltonian as 
\be 
\label{eq:polH2} 
\hat{H}_\text{pol}=\frac{1}{2}\Big(4\hat{\Pi}\hat{\phi}^{1/2}\hat{\Pi}+\hat{\phi}^{1/2}\Big), 
\ee
corresponding to the Schr\"odinger Hamiltonian \eqref{eq:schHordering} with $\alpha =1/2$ and $\beta = 0$. For reasons that will emerge shortly, we refer to \eqref{eq:polH2} as the nonsingular ordering. 

We write the basis states as $|\polv\rangle := |m\mmuu\rangle$, where $m \in \mathbb{Z}$ and $\mmuu := 2\mmu$. We then have 
\begin{align}
{\hat H}_{\mathrm{pol}} | m \mmuu \rangle 
= 
\frac{2}{\mmuu^{3/2}}
\biggl[
& 
\bigg( \big| m+{\textstyle\frac{1}{2}} \big|^{1/2} 
+ \big| m-{\textstyle \frac{1}{2}} \big|^{1/2} 
+ \frac{\mmuu^2}{4} {|m|}^{1/2} \bigg) 
 | m \mmuu \rangle 
\notag
\\
& 
- {\big| m+{\textstyle \frac{1}{2}} \big| }^{1/2}\, | (m+1) \mmuu \rangle 
- {\big| m-{\textstyle \frac{1}{2}} \big| }^{1/2}\, | (m-1) \mmuu \rangle 
\biggr].
\end{align}
The subspace in which $c_m = c_{-m}$ and the subspace in which $c_m = -c_{-m}$ are each invariant under the action of the Hamiltonian. These subspaces therefore decouple and it suffices to consider each of them individually. We refer to these subspaces respectively as the symmetric sector and the antisymmetric sector. 

The eigenvalue equation
\be 
\hat{H}_\text{pol} \sum_{m=-\infty}^{\infty} c_m|m\mmuu\rangle 
=M\sum_{m=-\infty}^{\infty} c_m|m\mmuu\rangle 
\ee
gives the recursion relation
\be 
\label{eq:rec} 
c_m\biggl[ \big| m+ {\textstyle \frac{1}{2}} \big|^{1/2}+ \big| m-{\textstyle \frac{1}{2}} \big|^{1/2}+
\frac{\mmuu^2}{4}|m|^{1/2}-\frac{\mmuu^{3/2}M}{2} \biggr]=
	\big|m-{\textstyle \frac{1}{2}} \big|^{1/2} c_{m-1}+\big| m+{\textstyle \frac{1}{2}} \big|^{1/2} c_{m+1}.
\ee
As the coefficients of $c_{m-1}$ and $c_{m+1}$ in \eqref{eq:rec} are nonvanishing for all~$m$, the recursion relation is nonsingular in the sense that the free initial data consist of $c_m$ at any two adjacent values of~$m$, and these initial data determine $c_m$ for all~$m$. Expecting a discrete spectrum, we look for solutions that are normalizable, 
$\sum_m {|c_m|}^2 < \infty$. Examination of the asymptotic form of \eqref{eq:rec} shows that such solutions have at $m\to\pm\infty$ the asymptotic form 
\begin{align}
\label{eq:approx_c} 
c_m = {|m|}^{\delta -1/4} \exp (\lambda\sqrt{|m|})
\Bigg[1+\frac{\mmuu^2}{8}+\sqrt{\Big(1+\frac{\mmuu^2}{8}\Big)^2-1}\,\Bigg]^{-|m|} \left[1+\mathcal{O}\bigl({|m|}^{-1/2}\bigr)\right],
\end{align}
where 
\begin{subequations} 
\label{eqs:deltalambda}
\begin{align}
\delta &= \frac{\mmuu^{3}M^2}{32}\times\frac{1+\dfrac{\mmuu^2}{8}}{\bigg[\bigg(1+\dfrac{\mmuu^2}{8}\bigg)^2-1\bigg]^{3/2}},\\
\lambda &= \frac{\mmuu^{3/2}M}{2}\times\frac{1}{\bigg[\bigg(1+\dfrac{\mmuu^2}{8}\bigg)^2-1\bigg]^{1/2}}. 
\end{align}
\end{subequations}

To find the mass eigenvalues numerically we use the shooting method. Given a trial value for~$M$, we choose an $m_0$ so large that $c_{m_0}$ and $c_{m_0-1}$ can be obtained from (\ref{eq:approx_c}) and iterate downwards using~(\ref{eq:rec}). In the symmetric sector we shoot for values of $M$ that give $c_1 = c_{-1}$, and in the antisymmetric sector we shoot for $c_0 =0$. The numerical accuracy is monitored by increasing $m_0$ until the results no longer change to the desired accuracy. 

Working to six significant figures, we find reliably the 10 lowest eigenvalues for $\mmu \lesssim 15$ and the 50 lowest eigenvalues for $\mmu \lesssim 2.5$. For small $\mmu$ the numerics becomes increasingly slow and we do not consider values of $\mmu$ below 0.005. 

\begin{figure}[hp!]
\begin{center}
\includegraphics[scale=0.3]{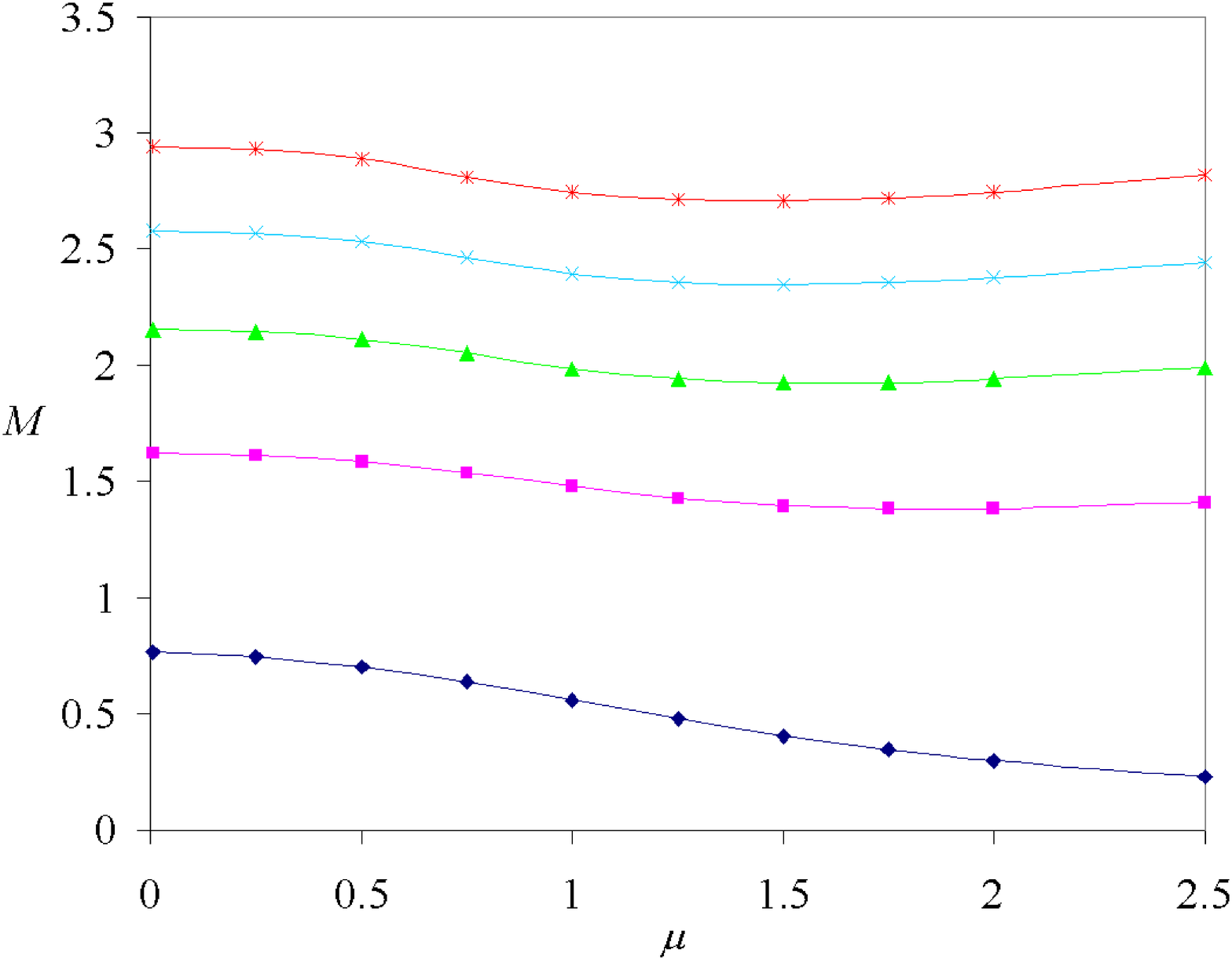}
\caption{Five lowest symmetric sector mass eigenvalues for $0.005 \leq \mmu \leq 2.5$. (The sizes of the markers do not indicate the numerical uncertainty.)} 
\label{fig:symmetric}
\end{center}
\hfill
\begin{center}
\includegraphics[scale=0.3]{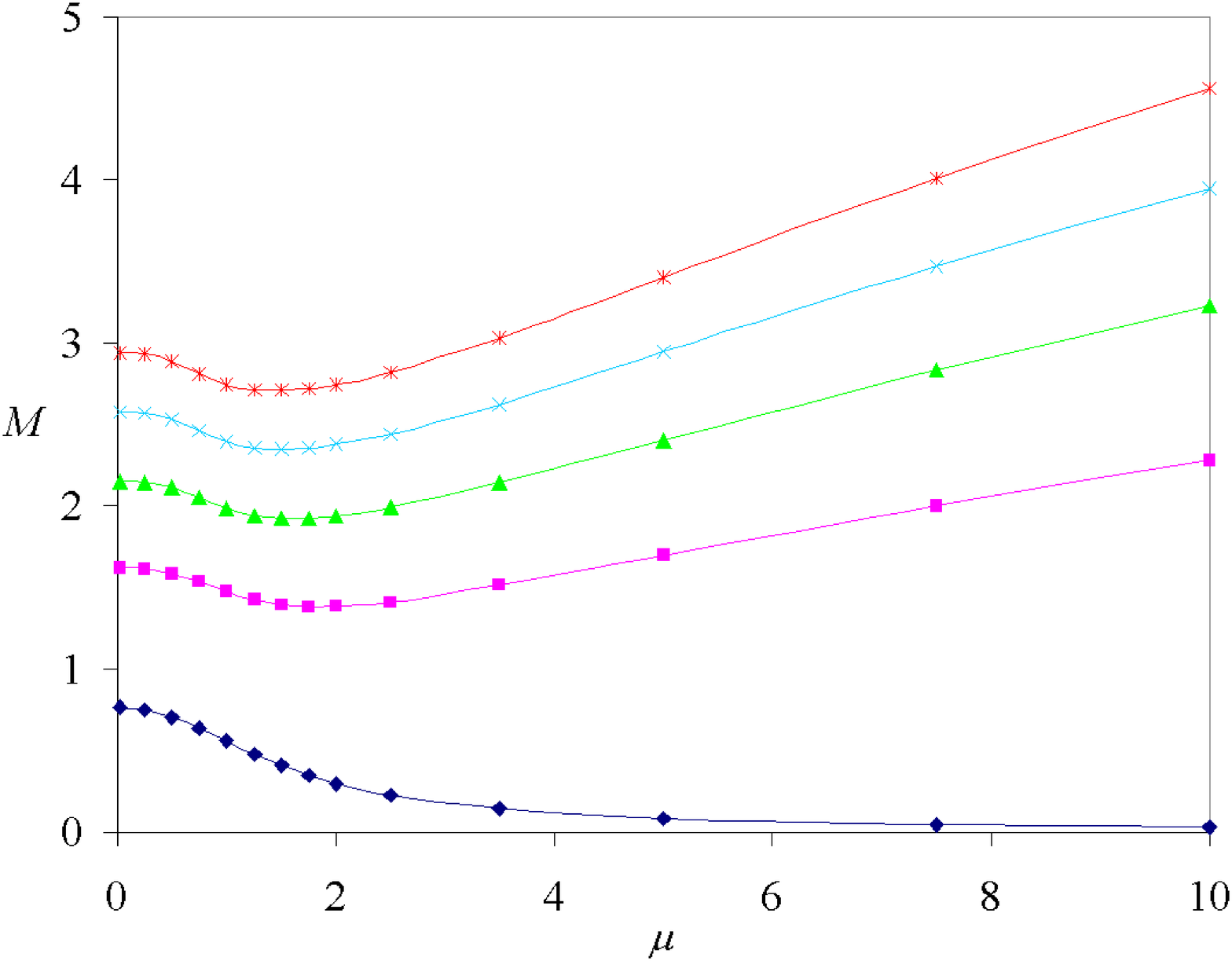}
\caption{Five lowest symmetric sector mass eigenvalues for $0.025 \leq \mmu \leq 10$.} 
\label{fig:symmetric2}
\end{center}
\end{figure}

\begin{figure}[hp!]
\begin{center}
\includegraphics[scale=0.3]{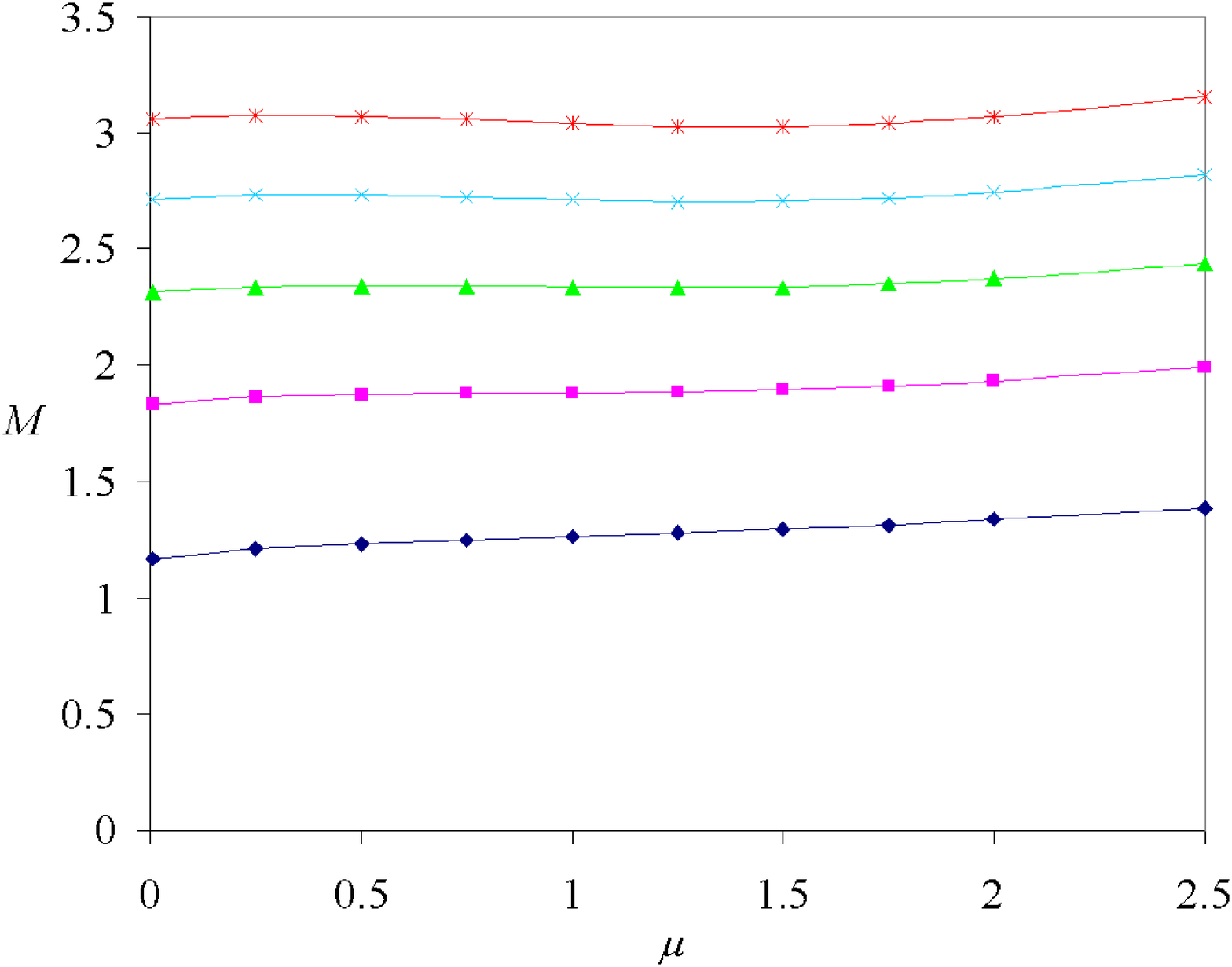}
\caption{Five lowest antisymmetric sector mass eigenvalues for $0.005 \leq \mmu \leq 2.5$.} 
\label{fig:antisymmetric}
\end{center}
\hfill
\begin{center}
\includegraphics[scale=0.3]{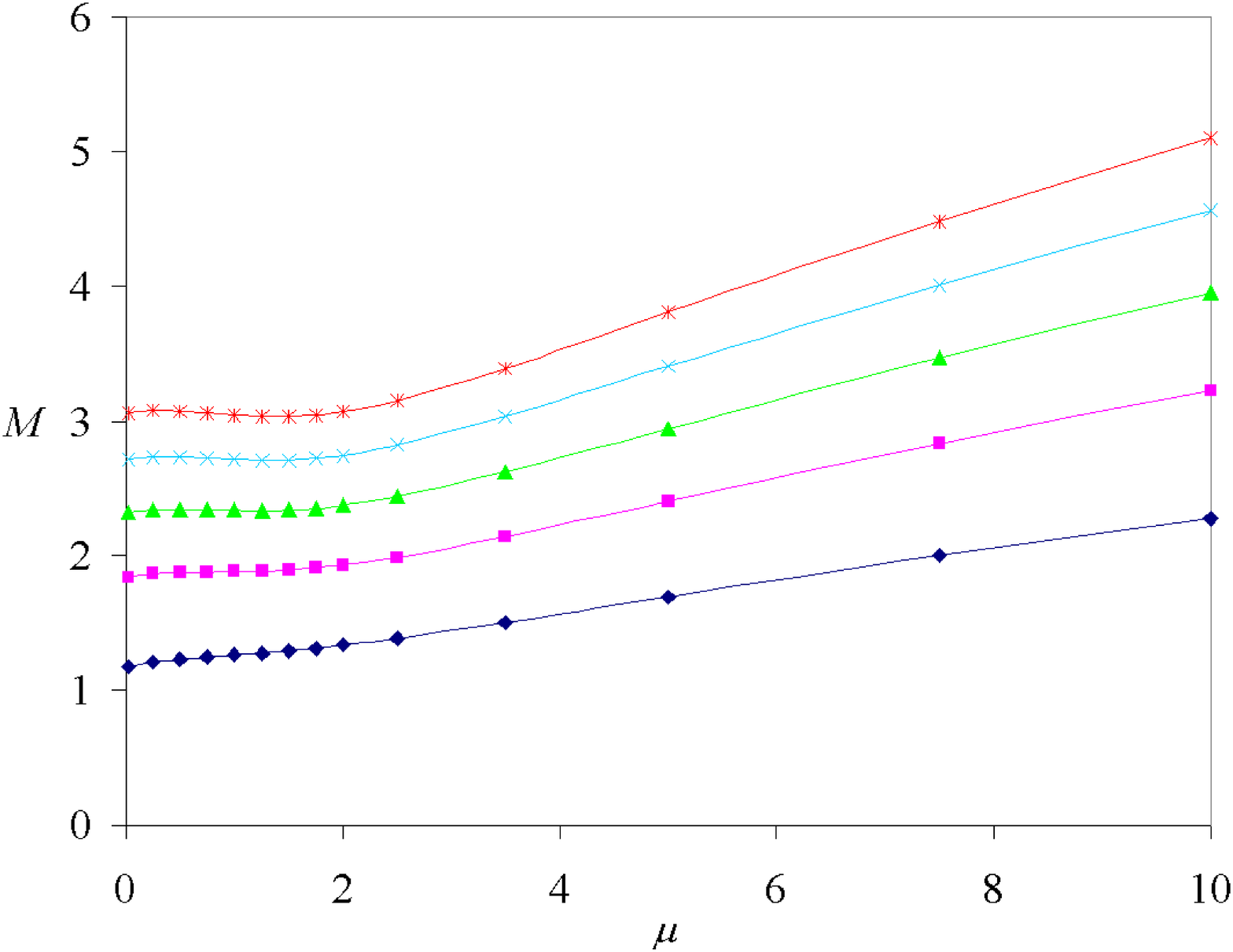}
\caption{Five lowest antisymmetric sector mass eigenvalues for $0.025 \leq \mmu \leq 10$.} 
\label{fig:antisymmetric2}
\end{center}
\end{figure}

Figures \ref{fig:symmetric} and \ref{fig:symmetric2} show the lowest five eigenvalues for $0.005 \leq \mmu \leq 10$ in the symmetric sector, and figures \ref{fig:antisymmetric} and \ref{fig:antisymmetric2} show the corresponding data in the antisymmetric sector. The ground state is in the symmetric sector, as expected, and the symmetric and antisymmetric sector eigenvalues alternate. The ground state eigenvalue decreases with increasing $\mmu$ and appears to asymptote to zero from above as $\mmu\to\infty$. The higher eigenvalues show a more complicated dependence on $\mmu$ for $\mmu \lesssim 2$, most of them having a minimum that is close to the minimum of the function $k(\mmu)$~\eqref{eq:kfunc-BS}, but they all increase in $\mmu$ for $\mmu>2$, at a rate that we will examine in Sec.~\ref{sec:semi}. The symmetric and antisymmetric sector eigenvalues do not appear to asymptote to each other as $\mu\to\infty$, but for fixed $\mmu$ the gap between the symmetric and antisymmetric sector eigenvalues gets smaller as the mass level increases. 

In the $\mmu \to0$ limit the eigenvalues appear to converge to definite limiting values. We wish to compare these limiting values to the eigenvalues in the corresponding Schr\"odinger theory, with the inner product \eqref{eq:ip-flat} and the Hamiltonian \eqref{eq:schHordering} with $\alpha = 1/2$ and $\beta = 0$. Mapping this Schr\"odinger theory to that of \eqref{eq:ipzero} and~\eqref{eq:Hzero}, discussed in~\cite{Louko:1996md}, we find that the parameter denoted in \cite{Louko:1996md} by $r$ has the value~$5/6$. The eigenfunctions are parabolic cylinder functions and the eigenvalue equation is (B10) of~\cite{Louko:1996md}. The small argument behaviour of the eigenfunctions 
is 
$\psi(\polv) = 
\cos(\theta) \polv^{1/2} 
\left[
1 + O(\polv^{3/2}) 
\right]
+ 
\sin(\theta)  
\left[
1 + O(\polv^{3/2}) 
\right]$, 
where $\theta \in [0, \pi)$ is the parameter that specifies the self-adjoint extension of the Hamiltonian. This small argument behaviour suggests that the symmetric and antisymmetric sector of the polymer theory should correspond respectively to the $\theta= \pi/2$ and $\theta=0$ Hamiltonians in the Schr\"odinger theory. With this identification of~$\theta$, we have verified that the polymer theory eigenvalues indeed converge to those of the corresponding Schr\"odinger theory, for the low eigenvalues where numerical accuracy allows us to test the convergence. The evidence for the three lowest eigenvalues in each sector is shown in Table~\ref{tab:evs}. 

\begin{table}[ht!]
\centering
\begin{tabular}{c|c c c}

\multicolumn{4}{l}{Nonsingular ordering, symmetric sector} \\[1ex]
\hline\hline
$\mmu  $ & $\qquad  M_0\quad $ & $\qquad M_1\quad$ & $\qquad M_2\quad$ \\
\hline
0.1 & 0.763255 & 1.62108 & 2.15328  \\
0.05 & 0.766485 & 1.62254 & 2.15461  \\
0.005 & 0.768132 & 1.62321 & 2.15517 \\
\hline
$\theta = \pi/2$ Schr\"odinger & 0.768184 & 1.62323 & 2.15518 \\
\hline\hline
\multicolumn{4}{l}{} \\[1ex]
\multicolumn{4}{l}{Nonsingular ordering, antisymmetric sector} \\[1ex]
\hline\hline
$\mmu  $ & $\qquad  M_0\quad $ & $\qquad M_1\quad$ & $\qquad M_2\quad$ \\
\hline
0.1 & 1.19141 & 1.85036 & 2.32984 \\
0.05 & 1.18178 & 1.84343 & 2.32411 \\
0.005 & 1.16602 & 1.83184 & 2.31424 \\
\hline
$\theta = 0$ Schr\"odinger & 1.15890 & 1.82661 & 2.30978 \\
\hline\hline
\multicolumn{4}{l}{} \\[1ex]
\multicolumn{4}{l}{Singular ordering} \\[1ex]
\hline\hline
$\mmu  $ & $\qquad  M_0\quad $ & $\qquad M_1\quad$ & $\qquad M_2\quad$ \\
\hline
0.1 & 1.31307 & 1.92381 & 2.38555 \\
0.05 & 1.31320 & 1.92415 & 2.38605 \\
0.005 & 1.31325 & 1.92427 & 2.38623 \\
\hline\hline
\end{tabular}
\caption{Three lowest eigenvalues in the symmetric and antisymmetric sectors 
with the nonsingular ordering and with the singular ordering as $\mmu\to0$. 
For the nonsingular ordering, the eigenvalues convergence to the corresponding Schr\"odinger eigenvalues.}
\label{tab:evs}
\end{table}

\subsection{Factor ordering with a singular eigenvalue equation}

In this subsection we order the  
polymer Hamiltonian as 
\begin{equation} 
\label{eq:polH1}
{\hat H}_{\mathrm{pol}} = \frac{1}{2} 
\left( 
4 {\hpolv}^{1/4} \, 
{{\hat\Pi}}^2 \, 
{\hpolv}^{1/4}
+ {\hpolv}^{1/2}
\right) , 
\end{equation}
corresponding to the Schr\"odinger Hamiltonian \eqref{eq:schHordering} with $\alpha =1/2$ and $\beta = 1/4$. For reasons that are about to emerge, we refer to \eqref{eq:polH1} as the singular ordering. 

Writing again $\mmuu =2\mmu$, we now have 
\begin{align}
{\hat H}_{\mathrm{pol}} | m \mmuu \rangle 
= 
\frac{2}{\mmuu^{3/2}}
\Biggl[
& 
\left( 2 + \frac{\mmuu^2}{4} \right) 
{|m|}^{1/2} | m \mmuu \rangle 
\notag
\\
& 
- {|m(m+1)|}^{1/4} | (m+1) \mmuu \rangle 
- {|m(m-1)|}^{1/4} | (m-1) \mmuu \rangle 
\Biggr].
\end{align}
The action of the Hamiltonian 
leaves the subspaces of positive $m$ and negative $m$ invariant and annihilates the subspace $m=0$. These three subspaces are thus entirely decoupled. The dynamics in the $m=0$ subspace is trivial, and the only energy eigenvalue is zero. The actions in the subspaces of positive and negative $m$ are isomorphic via $m \mapsto -m$, and it hence suffices to consider the subspace of positive~$m$. 

The eigenvalue equation 
\begin{equation}
{\hat H}_{\mathrm{pol}}
\sum_{m=1}^\infty c_m | m\mmuu \rangle 
= M \sum_{m=1}^\infty c_m | m\mmuu \rangle 
\end{equation}
gives the recursion relation 
\begin{align}
c_m 
\left[
\left( 2 + \frac{\mmuu^2}{4} \right) m^{1/2}
- \frac{\mmuu^{3/2}M}{2}
\right]
= 
m^{1/4}
\left[
{(m+1)}^{1/4} c_{m+1}
+ 
{(m-1)}^{1/4} c_{m-1}
\right] ,
\label{eq:sing-rec-c}
\end{align}
where $m = 1,2,3,\dots$. Note that this relation does not involve 
$c_0$ as its coefficient vanishes. The recursion relation is singular in the sense that evaluating \eqref{eq:sing-rec-c} for $m=1$ gives the condition 
\begin{align}
c_1
\left( 2 + \frac{\mmuu^2}{4} 
- \frac{\mmuu^{3/2}M}{2}
\right)
= 
{2}^{1/4} c_{2}, 
\label{eq:sing-condition}
\end{align}
and the free initial data hence consists of only~$c_1$. Examination of the asymptotic form of \eqref{eq:sing-rec-c} at $m\to\infty$ shows that the normalizable solutions have at $m\to\infty$ the asymptotic behaviour given by \eqref{eq:approx_c} and~\eqref{eqs:deltalambda}. 

The numerical problem is almost identical to that of Sec.~\ref{sec:nonsing}. We use \eqref{eq:sing-rec-c} to come down from a large~$m_0$, matching $c_{m_0}$ and $c_{m_0-1}$ to~\eqref{eq:approx_c}, and shoot for~\eqref{eq:sing-condition}.

\begin{figure}[hp!]
\begin{center}
\includegraphics[scale=0.3]{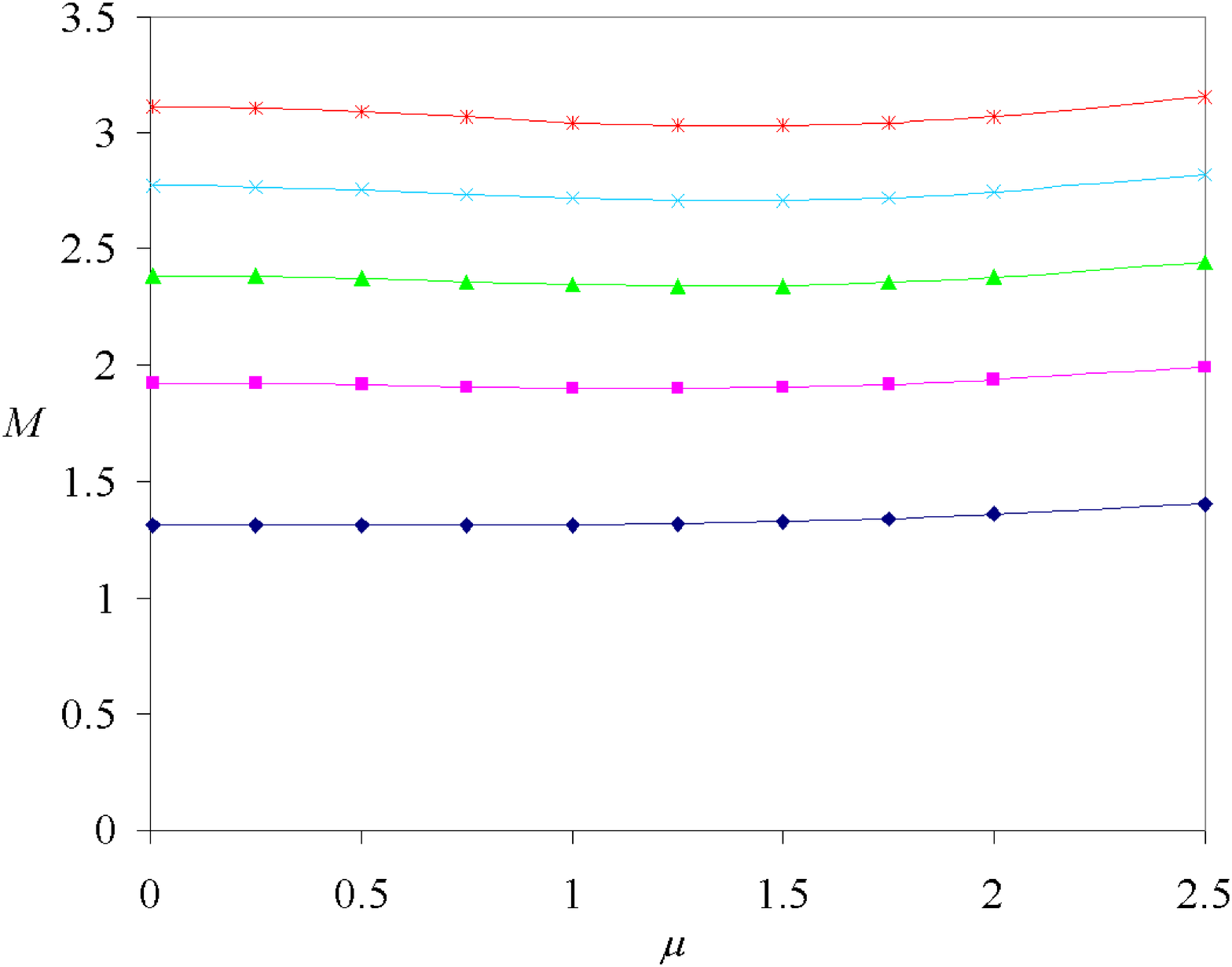}
\caption{Five lowest eigenvalues with the singular ordering for $0.005 \leq \mmu \leq 2.5$.} 
\label{fig:sing_asym}
\end{center}
\hfill
\begin{center}
\includegraphics[scale=0.3]{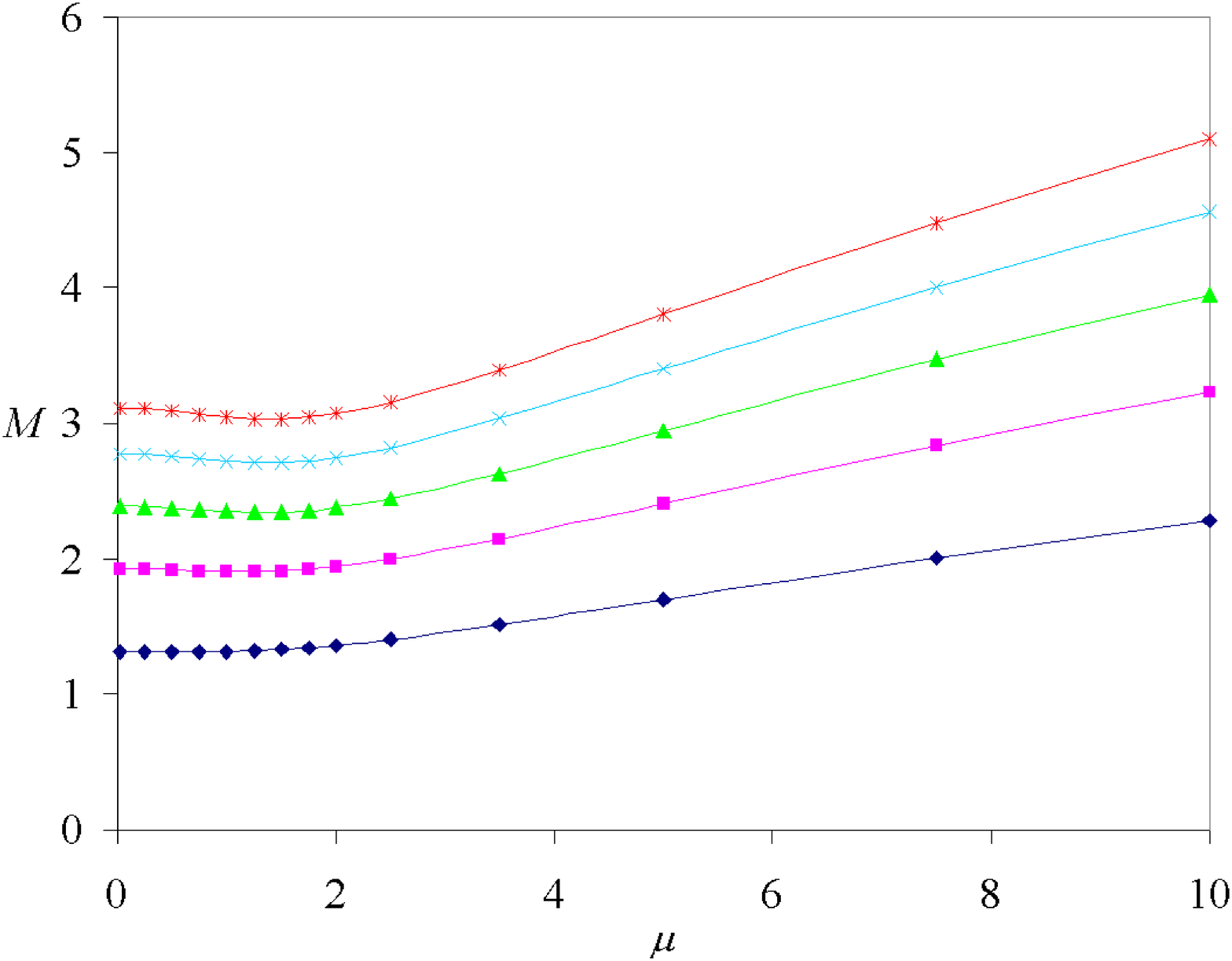}
\caption{Five lowest eigenvalues with the singular ordering for $0.025 \leq \mmu \leq 10$.} 
\label{fig:sing_asym2}
\end{center}
\end{figure}

\begin{figure}[hp!]
\begin{center}
\includegraphics[scale=0.3]{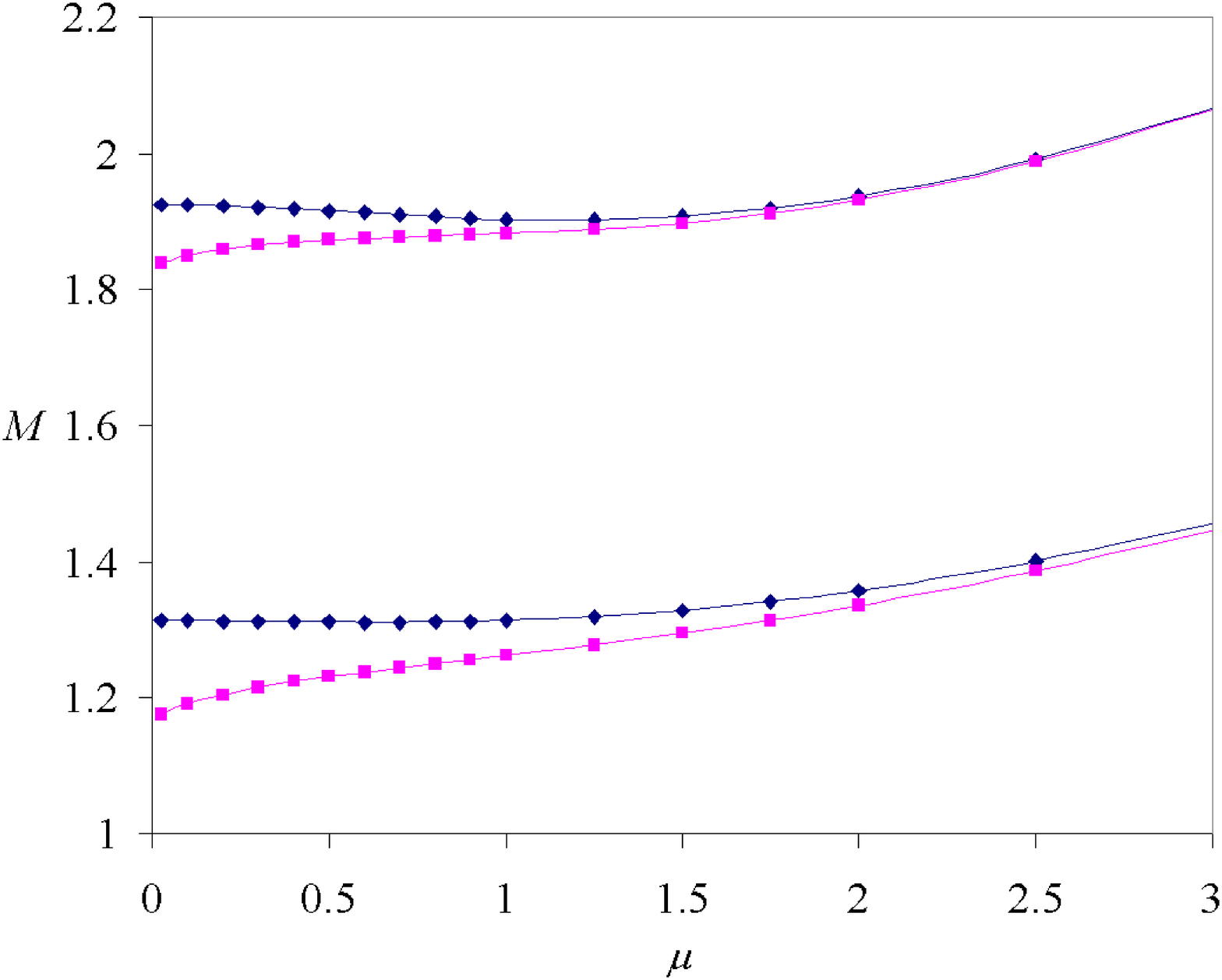}
\caption{Two lowest eigenvalues for the antisymmetric sector of the nonsingular ordering (purple curve with square markers) and 
for the singular ordering (dark blue curve with diamond markers).} 
\label{fig:asym_M0M1}
\end{center}
\hfill
\begin{center}
\includegraphics[scale=0.27]{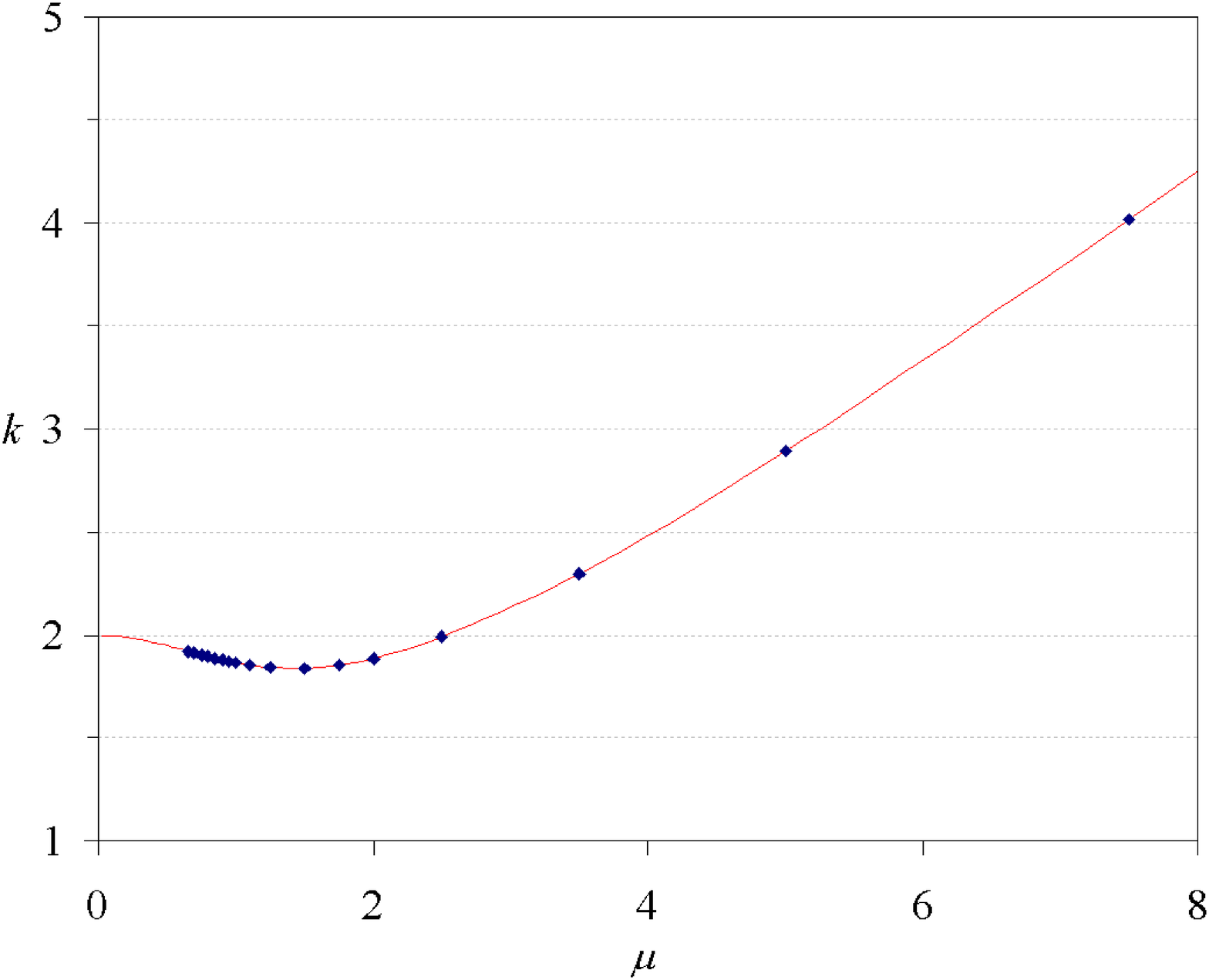}
\caption{The spacing factor $k(\mmu)$ as obtained from the numerical eigenvalues (discrete points) and from the Bohr-Sommerfeld estimate \eqref{eq:BS} (continuous curve).} 
\label{fig:k_mu2}
\end{center}
\end{figure}

Figures \ref{fig:sing_asym} and \ref{fig:sing_asym2} show the lowest five eigenvalues for $0.005 \leq \mmu \leq 10$. Comparison with Figs.\ \ref{fig:antisymmetric} and \ref{fig:antisymmetric2} shows that the eigenvalues are qualitatively very similar to those in the antisymmetric sector of the nonsingular ordering, albeit slightly higher: the two spectra appear to asymptote to each other as $\mmu\to\infty$, and the convergence gets more rapid as the mass level increases. At $\mmu\to0$, the eigenvalues appear to have well-defined limits that are genuinely above those of the antisymmetric sector of the nonsingular ordering, as illustrated in Fig.\ \ref{fig:asym_M0M1} and in Table~\ref{tab:evs}. 
We expect this limit to give the eigenvalues of the Schr\"odinger Hamiltonian \eqref{eq:schHordering} with $\alpha =1/2$ and $\beta = 1/4$, with an appropriate self-adjointness boundary condition, although we have not attempted to verify this. 

In summary, the spectrum of the singular ordering Hamiltonian \eqref{eq:polH1} is qualitatively very similar to the spectrum of the antisymmetric sector of the nonsingular ordering Hamiltonian~\eqref{eq:polH2}. The boundary condition \eqref{eq:sing-condition} that is enforced by the singular recursion relation \eqref{eq:sing-rec-c} plays the part of restricting the nonsingular recursion relation \eqref{eq:rec} to the antisymmetric sector.

\subsection{Semiclassical spectrum} 
\label{sec:semi}

We now turn to the semiclassical regime of the spectra. The Schr\"odinger quantization asymptotic estimate \eqref{eq:asym1} and the Bohr-Sommerfeld polymer result \eqref{eq:BS} suggest that the asymptotic behaviour of the large eigenvalues is  
\be 
\label{eq:linear} 
M_n^2 \sim k(\mmu)n + B,
\ee 
where $k(\mmu)$ is the Bohr-Sommerfeld spacing function \eqref{eq:kfunc-BS}, for both sectors of the nonsingular ordering and for the singular ordering. The additive constant $B$ could \emph{a priori\/} depend on the ordering, the sector and~$\mmu$. 

Our numerical results confirm this expectation in the range of $\mmu$ where the numerical accuracy is sufficiently good. The numerical limitations arise mainly from the computational noise that builds up after 50 or so eigenvalues, or for $\mmu\simeq 15$ already after 15 or so eigenvalues. For $0.65\leq \mmu \leq 15$ we find full agreement with~(\ref{eq:linear}). For $\mmu \lesssim0.65$ the noise piles up before the semiclassical regime is reached, with the notable exception of $\mmu\simeq 0.01$, where the spectrum appears to obey \eqref{eq:linear} already near the ground state with $k$ very close to~$2$. A~plot of the numerical results superposed on the Bohr-Sommerfeld $k(\mmu)$ is shown in Fig.~\ref{fig:k_mu2}. 

The numerical results give also information on the additive constant $B$ for $0.65 \leq\mmu\leq 15$. With the nonsingular ordering, we find that $B$ differs between the two sectors. In the symmetric sector $B$ is small and negative (taking the convention in which the ground state is $n=0$), and it is essentially independent of~$\mmu$. In the antisymmetric sector $B$ is positive, and it depends on $\mmu$ qualitatively in the same way as~$k(\mmu)$. With the singular ordering, $B$ is very close to that of the nonsingular ordering antisymmetric sector, and the difference appears to be decreasing as $\mmu$ increases (cf.\ the behaviour of the lowest two eigenvalues in Fig.~\ref{fig:asym_M0M1}). The difference is however small, and we have not been able to exclude the possibility that it might be an artefact of the numerics.

\section{Conclusions} 
\label{sec:conc}

In this paper we have addressed polymer quantization of spherically symmetric Einstein gravity, in a description in which the dynamical variables are associated with the Einstein-Rosen wormhole throat. Choosing to polymerize the area of the throat, we found numerically that the spectrum is discrete and positive definite, and the large eigenvalues imply an evenly-spaced area spectrum that is consistent with a Bohr-Sommerfeld semiclassical estimate. Issues of factor ordering arise in the kinetic term of the Hamiltonian, but the spectrum is largely insensitive to these issues except at the lowest few eigenvalues. In the limit of small polymerization scale the spectrum tends, within the numerical accuracy, to that of a conventional Schr\"odinger quantization. 

Our mathematical results are broadly similar to those obtained previously in polymer quantizations of the $1/r$ and $1/r^2$ potentials~\cite{hlw,klz}, despite the technical differences that arise from the singularities of these potentials. There is however a difference in the physical interpretation of the polymerization scale. The physical context of the $1/r$ potential is quantum gravity corrections to atomic physics, and in this case we expect the polymerization scale $\mmu$ to be at the Planck scale, which is much smaller than the Rydberg scale of the non-polymerized Hamiltonian. It was indeed found in \cite{hlw} that the polymer results are in this limit close to the usual Schr\"odinger results. The physical context of the $1/r^2$ potential is less clear, but as this system is classically scale invariant, the polymerization scale $\mmu$ affects the results only by setting the energy scale. Now, the physical context of our system is genuine gravitational dynamics, and one expects the polymerization scale $\mmu$ to be at the Planck scale, which is of order unity in our Planck units. Our numerical results for the spectrum appear reasonable for $\mu \approx 1$, but it is puzzling that the wormhole throat bounce in the effective polymer theory then occurs at a macroscopic scale, for a macroscopic black hole. It would be of interest to examine the motion of wave packets in the full polymer theory to ascertain that this macroscopic bounce is not an artefact of the effective polymer theory. 

The bounce necessarily occurs below the horizon, so the question of whether a macroscopic bounce scale has observational consequences for a galactic black hole, for example, depends on the details of the quantum corrected space-time exterior to the horizon. Reconstructing a quantum-corrected spacetime from the polymer throat theory would require additional input about the spatial hypersurfaces of the Hamiltonian foliation. It might be of interest to analyze this reconstruction in some geometrically simple foliations, such as the Novikov coordinates~\cite{MTW}, but the absence of local dynamical degrees of freedom would probably make it difficult to prefer any such foliation to another. The prospects of fixing a foliation by local criteria would be higher if our techniques could be generalized to systems that carry Hawking radiation, such as the spherically symmetric Einstein-scalar system~\cite{Ziprick:2009nd}. In the absence of such additional structure, however, one would hope that local changes in the foliation (at least those that preserve spherical symmetry) would not affect the physical predictions of the model. If this were not the case, it would reinstate in a much more serious way the problem of dependence on fiducial structures that occurs in the semi-classical polymerization of the homogeneous interior.

\section*{Acknowledgments} 

We thank Jonathan Ziprick for helpful discussions. 
GK was supported in part by the
Natural Sciences and Engineering Research Council of Canada. 
JL was supported in part by STFC (UK) grant PP/D507358/1.

\end{document}